# Colour changes upon cooling of Lepidoptera scales containing photonic nanoarchitectures

István Tamáska[1a], Krisztián Kertész[1b], Zofia Vértesy[1c], Zsolt Bálint[2d], András Kun[2e], Shen-Horn Yen[3f], László Péter Biró[1g]*

[1]Institute for Technical Physics and Materials Science, Research Centre for Natural Sciences, P.O.B. 49, H-1525 Budapest, Hungary
[2]Hungarian Natural History Museum, Baross utca 13, H-1088 Budapest, Hungary
[3]Laboratory of Natural Resource Conservation, Department of Biology and Institute of Life Science, National Sun Yat-Sen University, Kaohsiung, Taiwan, R. O. C.

**Abstract**
The effects produced by the condensation of water vapours from the ambient in the various intricate nanoarchitectures occurring in the wing scales of several Lepidoptera species were investigated by controlled cooling (from room temperature to -5 – -10 °C) combined with in situ measurement of changes in the reflectance spectra. It was determined that, due to this procedure, all photonic nanoarchitectures giving a reflectance maximum in the visible range and having an open nanostructure exhibited alteration of the position of the reflectance maximum associated with the photonic nanoarchitectures. The photonic nanoarchitectures with a closed structure exhibited little to no alteration in colour. Similarly, control specimens coloured by pigments did not exhibit a colour change under the same conditions. Hence, this effect can be used to identify species with open photonic nanoarchitectures in their scales. For certain species, an almost complete disappearance of the reflectance maximum was found. All specimens recovered their original colours following warming and drying. Cooling experiments using thin copper wires demonstrated that colour alterations could be limited to a millimetre, or below. Dried museum specimens do not exhibit colour changes when cooled in the absence of a heat sink due to the low heat capacity of the wings.

## Introduction

Animal colour is mostly due to the spectrally selective reflection of incident light. It can be associated with absorbing pigments (Zollinger et al. 2003; Nijhout 1991), which are often referred to as chemical colour, and/or selective reflection caused by structural properties, which is usually referred to as physical or structural colour (Joannopoulos et al. 2008; Biró and Vigneron 2011). Some physical colours can be produced by photonic crystals (Seago et al. 2009; Wilts et al. 2009; Shawkey et al. 2009a; Poladian et al. 2009; Ghiradella and Butler 2009), which are periodic dielectric structures in space composed of two or more media with different optical properties. These structures have a forbidden gap, within which photons with certain energies cannot propagate and are completely reflected by the surface of the photonic crystal. Since 1987, when Yablonovitch (Yablonovitch 1987) and John (John 1987) described structures (i.e., nanoarchitectures) with photonic band gaps, the field has been developing at an increasingly rapid pace.

An enormous variety of photonic structures can be found in nature. Butterflies, beetles (Srinivasarao 1999, Seago et al. 2009, Shawkey et al. 2009a, Ghiradella & Butler 2009, Biró et al. 2010) and even plants (Vigneron et al. 2005a, Glover & Whitney 2010) exhibit these kinds of structures. It has recently been shown that, in certain species, such as the sexually dichromatic species *Hypolimnas bolina*, females prefer conspecific males that possess bright iridescent blue/ultraviolet dorsal ornamentation (Kemp 2007) so that the "quality" of the photonic nanoarchitectures is well preserved from generation to generation. As emphasised by several recent reports, visual signals from conspecific individuals make important contributions to the mating behaviours of butterflies (Fordyce et al. 2002, Oliver et al. 2002, Bálint et al. *in press*).

Several recent reviews have been published on photonic nanostructures of biological origin (Parker 2002, Vukusic & Sambles 2003, Fratzl 2007, Kinoshita 2008, Biró & Vigneron 2011). Butterflies are diverse and highly practical examples of these structures, as their colour is generated by pigments and by various nanostructures (Kertész et al. 2006a, Kinoshita & Yoshioka 2005, Vértesy et al. 2006, Welch & Vigneron 2007, Vukusic et al. 1999, Vukusic & Stavenga 2009, Argyros et al. 2002). The essentially flat wings of butterflies can be easily processed into convenient samples for both scientific examination and practical applications, such as solar cells (Zhang et al. 2009), sensors (Potyrailo et al. 2007, Biró et al. 2008) and many other uses (Huang et al. 2006, Berthier et al. 2007).

Lepidopteran wings and the scales covering them have a complex geometric structure on macroscopic- and microscopic-length scales. The membranes of the wings are generally covered with partially overlapping scales on both sides. Typical scale dimensions are in the ranges of 50 – 100 μm for length and 15 – 50 μm for width. Most species have two distinct layers of different scales: ground scales and cover scales. While the underside of the scale is rather featureless, the externally visible top surface usually exhibits a complex structure from the micron to the nanometre range. This top surface or the volume of the scale contains photonic nanostructures if the colour of the wings is of structural origin. These nanostructures are mainly constructed of a chitinous matrix, including air holes (Scoble 1995, Kristensen 2003, Berthier 2006). These biological structures attract more and more attention because they constitute a transition between random and crystalline order (Shawkey et al. 2009; Liu et al. 2011). Recent papers reported that even a complete photonic band gap can be found in such partially ordered materials (Florescu et al. 2009).

It often may not be a straightforward task to determine if a certain butterfly exhibits structural colour, especially in the case of species presenting non-iridescent, matte structural colour, as is seen on the ventral side of *Cyanophrys remus* (Kertész et al. 2006b) or *Callophrys rubi* (Ghiradella & Radigan 1976). Of course, transmission and scanning electron microscopy (TEM and SEM, respectively) can reveal structural details, but these methods require time-consuming sample preparation (Shawkey et al. 2009b), expensive equipment and physical destruction of the (potentially rare) exemplars. Sophisticated optical characterisation (Kertész et al. 2006b, Wilts et al. 2009) can also reveal the presence of photonic nanoarchitectures, but the required instruments and the expertise may not be available in all laboratories; again, destructive sample preparation steps usually cannot be avoided. The method described by Mason (Mason 1923), based on soaking the specimen (sometimes for days) in liquids with a refractive index matching the refractive index of the nanoarchitecture, may also be an alternative. Dropping liquids (e.g., oil or ethanol) onto wings could identify nanostructures, but because these liquids cannot be found in the natural environments of butterflies, this method may raise concern from museum curators that it could damage the valuable samples. This type of method may induce persistent changes in the status of a specimen.

In this paper, we present a simple method based on cooling in ambient air to observe and investigate the colour-change of the wings of several Lepidoptera species coloured by photonic nanoarchitectures in detail. Cooling the butterflies and allowing water vapour to condense quickly onto the micro- and nanostructures induces colour change in their scales, which possess structural colour arising from open nanoarchitectures. This simple method allows the observer to determine which colours are of structural origin, as pigment-coloured wings do not show the colour change observed in the case of structural colours. During the cooling experiments, we identified significant differences in a few species in the time lapse of the colour change and how it changed; these differences originate from differences in air circulation in different nanoarchitectures. We also attempted to highlight some basic ideas that underlie the interaction of water with wings and the colour-generating nanoarchitectures. As softening in water vapour is the standard procedure for setting butterflies, water vapour condensation should not raise concerns from museum curators. It is well known that most Lepidoptera wings are hydrophobic (Wagner et al. 1996), which prevents liquid water from penetrating into the photonic nanostructure because water droplets roll off the surface (Wagner et al. 1996), thus preventing the soaking of the wings in rain. Using the cooling method described herein, this difficulty can be easily avoided.

## Materials and Methods

*Eterusia taiwana* (Lepidoptera: Zygaenidae) moths were obtained from the collections of the Laboratory of Natural Resource Conservation, Department of Biology and Institute of Life Science, National Sun Yat-Sen University, Taiwan. All other samples were obtained from the Lepidoptera collection of the Hungarian Natural History Museum.

For structural investigation of the wings of Lepidoptera species, scanning electron microscopy (SEM) and transmission electron microscopy (TEM) were used. SEM samples were prepared by cutting off pieces of the wings, which were then attached to stubs with double-sided conductive tape. All samples were subsequently coated with 15 nm sputtered gold to allow for examination without charging. Cross-sectional TEM samples were prepared by embedding pieces of wings in special resin. Thin sections with a thickness of 70 nm were cut using an ultramicrotome and were transferred to copper grids.

For optical characterisation, reflectance measurements were performed with an Avaspec 2048/2 fibre-optic spectrometer. A coaxial illuminator/pick-up fibre was used under normal incidence conditions. Measurements were performed with unpolarised light using a combined halogen-deuterium light source. An Avaspec diffuse white standard was used as a reference sample. This system allows accurate measurements in the 200 – 1000 nm wavelength range.

Measurements performed under controlled cooling were combined with spectroscopic observation of the colour change. Three types of cooling apparatuses were used: (i) a custom-built Peltier cooler, which could be placed under the combined illuminator/pick-up fibre of the spectrometer; (ii) the deep freezer compartment of a commercially available refrigerator; and (iii) thin copper wires frozen into ice blocks within the deep freezer compartment. Peltier elements use the thermoelectric effect to directly convert electric power to a temperature difference. They are usually thin plates (3 mm in our experiments) and have two large sides (squares with 5 cm sides in our experiments) where the temperature difference can be created. Two Peltier elements in series were mounted on a CPU cooler fan for better heat dissipation. A shield was mounted around the Peltier elements to direct the airflow away from the samples and to ensure a uniform temperature. With this construction, a temperature of -5 ºC could be achieved in a normal room environment. Spectroscopic measurements were taken every second during the controlled cooling of the samples. An entire cooling cycle was typically completed over a time range of minutes. While Peltier coolers are not expected to be present in all laboratories working with butterflies, refrigerators are common in most of them. We performed the experiments using the deep freezer compartment, cooling the butterflies for approximately 15 minutes.

## Results

### Colour change in wings due to cooling

The initial motivation for our work was provided by the experimental observation of colour changes in living *E. taiwana* moths. The moths were kept in a refrigerator at approximately 5 °C to extend their lifespan. The cooled living *E. taiwana* moths exhibit colour changes from green to dark brown to almost black when removed from the refrigerator. After being maintained at room temperature for a few hours, the moth regained its former colouration. The experiment was repeated with dried museum samples, but no colour changes were observed. When cooling the dried wing pieces that had been removed from the bodies using the Peltier cooler, colour changes could be easily observed (Fig. 1). The experiment was repeated under an optical microscope. Water condensation (small water drops) and freezing (ice) could be clearly observed on the surface of the Peltier cooler.

Further experiments were performed with wing samples from other Lepidoptera species with photonic nanostructures in their scales that were cooled with Peltier elements [from Lycaenidae: *Albulina metallica* (blue dorsal wing surface), *Callophrys rubi* (green ventral wing surface), *Cyanophrys remus* (blue dorsal and green ventral wing surfaces) and *Polyommatus daphnis* (blue dorsal wing surface); and from Nymphalidae: *Morpho aega* (blue dorsal wing surface)]. As the width of the Peltier element is 5 cm, only small pieces of lepidopteran wings were used to allow simultaneous cooling of several samples. Colour changes could be easily observed on every sample (Fig. 2).

A simpler cooling method involves the use of a household refrigerator (cooling temperature: -10 °C). Whole specimens can be cooled with an individual heat sink (glass microscope slides) mounted under each wing, allowing non-destructive measurements to be performed. Care has to be taken to provide good thermal contact between the microscope slide and the wing; a thin wire will suffice for this purpose, as shown in Fig. 3. Four butterflies were mounted on a large substrate with glass plates under each wing. The samples were cooled for 30 minutes in the refrigerator before being removed to an ambient room environment. The necessity of the heat sinks can be observed on butterflies in Fig. 3 a. and c. Colour changes cannot be observed near their bodies, where the wings are rapidly heated due to the absence of a cold thermal sink. As lepidopteran wings are not rigorously flat, some parts of the wings that are slightly closer to the cooled surface exhibit an earlier colour change than others. A sample of the butterfly *Lycaena virgaureae* was used for comparison, as it is not coloured by photonic nanoarchitectures, but by pigment (Vigneron et al. 2005b). It did not exhibit any colour change upon cooling.

**Spectroscopic measurements**

*E. taiwana* wing samples were mounted on the surface of the top Peltier element, and spectroscopic measurements were performed in situ during cooling (Fig. 4). The measurements began from a dry status (curve 1 in Fig. 4). After applying the previously determined voltage to the Peltier elements, the temperature dropped below the dewpoint in a few minutes (approx. 10 °C). When the dewpoint was reached and condensation of the water vapour began, the reflectance maximum shifted relatively rapidly to longer wavelengths (curve 2 in Fig. 4), and the green/blue moths became dark-brown/green. The samples were maintained at that temperature range for approximately 5 minutes after the shift, but the reflectance spectrum did not change. Then, the temperature was set to below 0 °C, and the wing pieces became completely frozen. After for waiting 5 minutes during which the position of the reflectance maximum did not change (curve 3 in Fig. 4), the applied voltage was turned off to allow the samples heat up to ambient temperature (23 °C). When melting began, the reflectance maximum shifted to shorter wavelengths with rapid progress (from curve 3 to curve 4 in Fig. 4) and stopped for approximately one minute when the ice in the wings was completely melted (0 °C). In a similar process, when the water in the samples began to evaporate, the reflectance maximum shifted towards shorter wavelengths and increased further in intensity until a colour close to the original colour was achieved. Full restoration of the wing colour was accomplished by drying overnight, indicating that a certain small but measurable fraction of water penetrated into the chitinous material. Similar spectroscopic measurements were performed on several other lepidopteran species, such as the well-characterised species *C. remus* (Kertész et al. 2006b) (Fig. 5). During cooling, this green species became brown. Additionally, the temperature of the Peltier cooler and the shift of the reflection maximum can be seen for the green ventral side of *A. metallica* in Fig. 1 of the Supplementary Material.

Further experiments were performed on several species to investigate the time dependence of the colour change (see Fig. 2 of the Supplementary Material for an example). The samples were mounted on the Peltier cooler and cooled to approximately 10 °C, the temperature at which condensation can begin. After 300 – 400 seconds, the cooling apparatus was turned off, and the samples were allowed to warm up to ambient temperature. The time-dependent shift of the reflectance peaks was investigated. The resulting curves depend on the complex structural differences between the wing scale nanoarchitectures and vapour condensation rates. Further investigation with more precise control of environmental conditions will be required to obtain additional insight into the processes taking place.

Repeated cooling – heating cycles were used to examine the long-term behaviour of the spectral shift. The samples were cooled (approx. 10 °C), and after the reflectance maximum had not changed for 5 minutes, the samples were heated to ambient temperature (23 °C); following an additional 5 minutes, the whole cycle was repeated. After approximately the third cycle, the butterflies completely lost their colour. Furthermore, chitin absorbed water, and the wings became malleable. Drying the samples for a few days fully restored the initial conditions of the wings. Observation of wing softening is not surprising, as this is the common preparation procedure for fixing the position of museum specimens.

**Local cooling by heat sink**

This cooling method can also be used to change the colour of lepidopteran wings locally. Cooling wings in specified small areas makes it possible to "write" on the wing surface. One method for this involves the use of a matrix built from miniature Peltier cells (each cell is used as a pixel), but a simpler method involves using cold wires below the wings. A cold wire cools a small area around it. If the temperature around the wire is lower than the dewpoint, condensation begins and changes the colour of the wing.

In our experiments, one end of thin copper wires was frozen into ice blocks, while the other end was used to form horizontal characters above the ice blocks. The butterfly *C. rubi* was placed onto these characters, and the characters were shaped to provide the best thermal contact with the wing membrane (Fig. 6). Copper has very good heat conductivity; it can easily transfer heat from the wings to the ice blocks, keeping the area cool around the wires on the wings. The temperature in the cooled area can be set by adjusting the distance between the ice and the characters. If the temperature of the wires is sufficiently high to be close to the dewpoint, the coloured area around them will be narrower. If the wing is too close to the ice block, the temperature of the entire wing can be lower than the dewpoint; this causes the entire wing to change colour. The coloured area around the wires widens in time as water vapour condenses. The writing begins to appear a few seconds after placing the butterfly onto the wires, and it can be clearly observed after waiting another minute.

## Discussion

### Role of the heat sink in the colour change

The difference between the dried and living *E. taiwana* moths in our first observations is that the body of the living specimens has a higher heat capacity due to its water content, and it is able to act as a heat sink. In a resting position, these moths keep their closed wings near their bodies, unlike many butterflies. The dried museum specimens have their wings positioned on the sides of their bodies, as is usual for museum exemplars. Additionally, they have lost most of the water content present in the living insect. Thus, they have a low heat capacity, so they cannot provide an adequate heat sink; the wings are heated to room temperature before significant vapour condensation could take place. This might be the reason that these specimens do not exhibit any colour change after being removed from the deep freezer compartment of the refrigerator (cooling temperature: -10 °C). If a large heat sink is provided by mounting microscope slides under the wings to increase the heat capacity of the combined system (wing + microscope slide), the wings can be kept cool for a sufficient amount of time to allow condensation to take place.

Investigation of the possible biological significance of the colour changes of wings (when cooled below the dewpoint in an ambient atmosphere) with respect to the survival strategy of these species might be a very interesting subject for visual ecology, but it is outside the scope of this paper.

### Water penetration and wing softening

As chitin absorbs water, it swells, thus altering the thickness of small structures made of chitin. This swelling effect can be observed in Figs. 4 and 5, where the reflection curves do not immediately return to the initial position, and the maxima have lower intensities after the cooling-heating cycle. However, overnight drying will fully recover the initial shape and position of the reflection curves. Because total swelling of the structures can be achieved by several cooling cycles (usually more than three), it is important that the reflection variations of several species are compared in the first cooling cycle to minimise the swelling effect.

When preparing museum specimens, butterflies are usually softened in saturated water vapours for hours or days to cause them to become malleable so they can be set. This is a slow process, as water vapours have to penetrate into the volume of the specimen aided by daily temperature variations. Although a sample absorbs significant amounts of water, in most cases, no liquid water will be present on its surface. A relaxed status of the wings is achieved from the repeated cooling-heating cycles within a few minutes instead of the days required for saturated water vapours.

### Colour change and intensity variations in spectroscopic measurements

The photonic crystal structures in lepidopteran wings are mainly constructed from chitin and air. Chitin has a moderate refractive index (n =1.56) (Vukusic et al. 1999). The position of the reflectance maximum ($\lambda_{max}$) can be easily estimated with a simple equation: $\lambda_{max}=n_{eff}*d/m$ (Vigneron et al. 2006), where $n_{eff}$ is the average refractive index; d is the thickness of one period in a periodic structure; and m is an integer. When air (n = 1) is replaced with water (n = 1.33), the contrast of the refractive indices decreases, which in turn will modify the average refractive index value. This decrease of the refractive index contrast shifts the stop-band (i.e., the reflectance maximum) to higher wavelengths, as can be observed in the spectral measurements presented in Figs. 4 and 5 between the curves labelled 1 and 2 The reverse process can also be observed between the curves labelled 4 and 5 Freezing the samples generates quite different results. For *C. remus,* the position of the reflectance maximum is changed only moderately by freezing, while for *E. taiwana,* a clear spectral shift occurs. Structural differences are likely responsible for these different behaviours, as will be discussed below in the context of the SEM figures.

The change in the intensity of the reflectance maximum during the cooling experiments depends on various parameters, such as the modifications induced in the surface geometry of the wing by wetting and freezing, the deformation of the chitin structures due to water absorption and the presence of water and ice on top of the scales. All of these parameters can differ from individual to individual. Due to this complex behaviour, a precise interpretation of the observed variations in intensity is beyond the scope of this paper.

**Effect of the nanoarchitecture during cooling**

During the simultaneous cooling experiments with the Peltier cooler (Fig. 2), it was observed that the wings of different Lepidoptera species exhibited slight differences in the time elapsed from the initiation of cooling until colour changes were clearly observed. This can be explained by taking into account the differences in the colour-generating nanoarchitectures in the scales.

If the surface temperature drops below the dew point, condensation begins on the surface, and it is quite difficult to calculate the exact condensation rate. The diffusion and heat equations have to be solved in parallel, making the boundary conditions match (for example, see Kandlikar 1999). We will not attempt to solve this problem, but we will attempt to highlight some essential features of the process. Lepidoptera wings are quite complex structures, possessing a wing membrane and usually several layers of scales. Colour is measured in the upper part of the wing, and cooling is applied to the lower part. The transportation of heat from the upper scales to the lower scales is quite complex. Therefore, the interactions between different scales and between the scales and the membrane have to be taken into account. As the observation of colour change occurs on the upper part of the wing, we attempt to restrict our interpretation to this part. When water vapour begins to condense, it will decrease the relative humidity below the critical value; for further condensation, a new supply of wet air is required. The more open the structure is, the more easily the supply of wet air will penetrate it. The fastest colour changes were observed for *C. remus*, *P. daphnis* and *C. rubi*, which have similar nanoarchitectures, while slightly slower colour modifications were observed for *A. metallica* and *E. taiwana*. The scales of *E. taiwana* exhibit a quite different, more closed structure compared to the other species examined in this work; the surfaces of their scales have fewer and smaller holes than the scales of the investigated butterflies (Fig. 7). It is possible that in closed nanoarchitectures, such as those of *Chrysiridia ripheus*, the supply of wet air will be insufficient for colour changes to occur (Fig. 7d). The cooling experiments demonstrated this to be the case; no significant colour change could be observed in this species.

## Conclusions

Our investigation of the rapid cooling of butterfly and moth wings in ambient air demonstrated that the wings of these species, which have photonic crystal-type nanoarchitectures with open structures, exhibit clear changes in coloration. Our data show that the durations and magnitudes of the colour changes and the ways in which they occurred were specific for certain structures (i.e., species), but small variations may also occur from individual to individual. In particular, in some closed photonic nanoarchitectures, such as those of the moth *C. ripheus,* a rapid colour change is not observed because water vapour cannot easily penetrate such closed structures. Therefore, if colour change is observed on cooling, this is a clear indication that the scales of the examined butterfly and moth species are coloured by photonic nanoarchitectures, while the absence of the colour change cannot be interpreted as unambiguous evidence that the scales are coloured only by pigments. In these cases, detailed structural (i.e., SEM and TEM) and spectral investigation may be required. Nevertheless, this quick and simple test can prove very useful in the rapid screening of species to determine whether it may be worth performing the time-consuming and expensive high-resolution characterisation procedures. At the same time, this screen provides information regarding the open (*C. remus, C. rubi*) or more closed (*E. taiwana, A. metallica, C. ripheus*) characters of the structures generating the structural colours.

Cooling can be achieved using the deep freezer compartment of a common refrigerator if a large heat capacity heat sink is mounted in good thermal contact with the wings. This makes large-scale application of the method possible. It is worth pointing out that well-dried collection specimens will not exhibit a colour change without an appropriate heat sink under their wings. Another important advantage of cooling in the refrigerator may be that it is non-destructive; even rare exemplars could be examined in this way without risking damage to the samples.

If repeated cooling and heating cycles are performed in succession, the wings can be softened in minutes. This rapid softening is attributed to repeated water vapour condensation taking place in the interior of the micro- and nanostructures of the scales and of the wing membranes. This will render the wings malleable without destruction of the exemplars or the need to keep the butterflies in saturated water vapours for several days.

If local cooling is provided, such as by thin copper wires, it is possible to "write" on wings possessing structural colours. This colour change effect upon water vapour condensation could be used to construct a tuneable photonic crystal-based flat panel display. The colour change can be induced by the local condensation of water vapours, which are always present in ambient air. Each pixel should be mounted on a miniature Peltier element, and the temperature (and colour) could be set with active feedback. The similar process of capillary condensation has been employed for optical switching in porous optical superlattices (Barthelemy et al. 2007) and is used by the beetle *Dynastes hercules* to reversibly switch colour depending on the relative humidity of the environment (Rassart et al. 2008).

## Acknowledgements

The work in Hungary was supported by OTKA grant PD83483. The collaboration of Hungarian and Taiwanese scientists was made possible by a joint agreement between the Hungarian Academy of Science and the NSC in Taiwan for mobility. K. Kertész gratefully acknowledges financial support from the János Bolyai Research Scholarship of the Hungarian Academy of Sciences.

**FIGURE CAPTIONS**

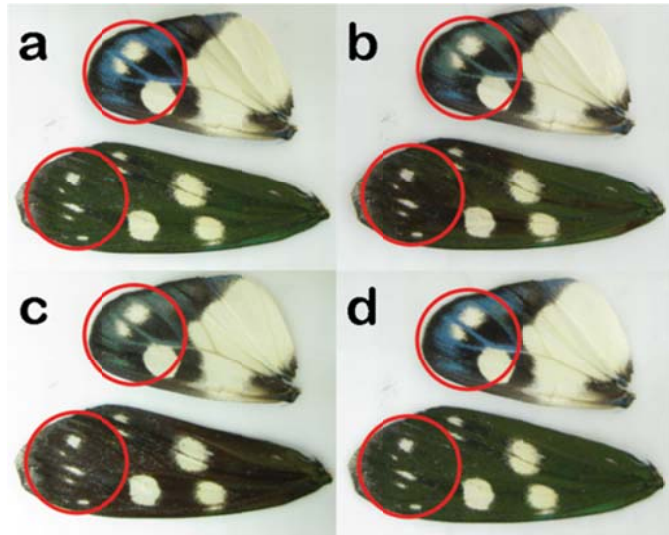

Figure 1: Colour change on the wings of *Eterusia taiwana* due to water vapour condensation and freezing caused by cooling. Blue-black-white dorsal hindwing surface and green-white dorsal forewing surface. a) At room temperature (22 °C); b) after vapour condensation onto the structures (10 °C); c) after freezing of the wings (-5 °C); d) after cooling and reheating to room temperature, when all of the water had evaporated from the wings.

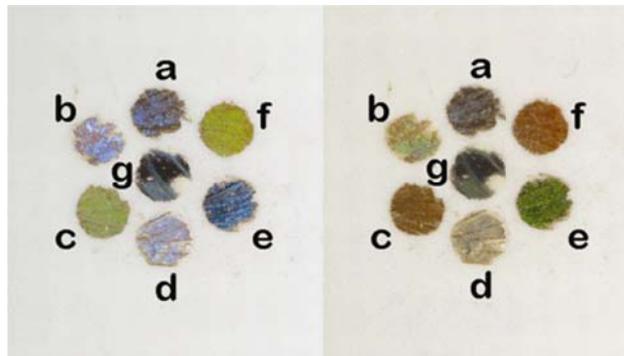

Figure 2: A simple display consisting of pieces of Lepidoptera wings before cooling (left) and after cooling (right). Colour changes can be observed due to water vapour condensation onto the photonic structures of the scales. The pieces of the wings are from: a) *Albulina metallica* (blue dorsal wing surface); b) *Morpho aega* (blue dorsal wing); c) *Cyanophrys remus* (green ventral wing); d) *Polyommatus daphnis* (blue dorsal wing); e) *Cyanophrys remus* (blue dorsal wing); f) *Callophrys rubi* (green ventral wing); and g) *Eterusia taiwana* (blue dorsal wing).

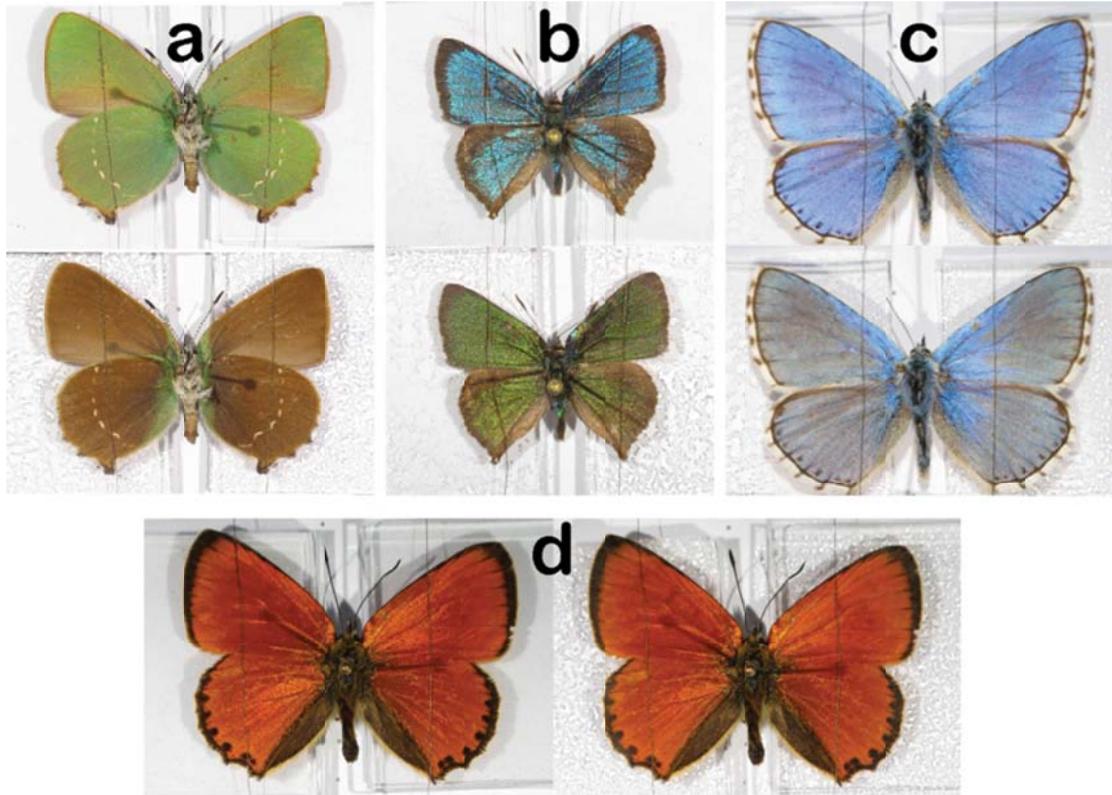

Figure 3: Lepidoptera species cooled in the refrigerator. a), b), c) Species with photonic structures in their scales and d) a species without photonic structures in its scales. Colour changes can be seen for species with photonic nanostructures, and the necessity of the heat sink can be observed near the bodies of a) and c), where the colour did not change. Under identical conditions, the butterfly coloured by pigments shown in d) did not change colour. a) *Callophrys rubi*, b) *Cyanophrys remus*, c) *Polyommatus bellargus*, d) *Lycaena virgaureae*.

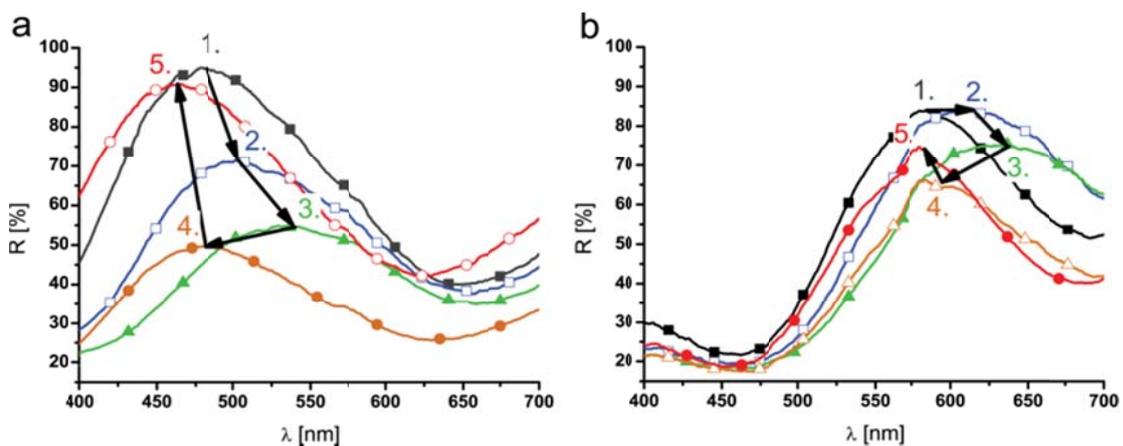

Figure 4: Spectral measurements on the blue/dorsal a) and green/ventral b) surfaces of the wings of the moth *Eterusia taiwana*. The numbered curves show the stationary reflectance curves in various cooling stages: 1. dry status (23 °C); 2. condensation (10 °C); 3. freezing (-5 °C); 4. melting (0 °C); and 5. evaporation processes.

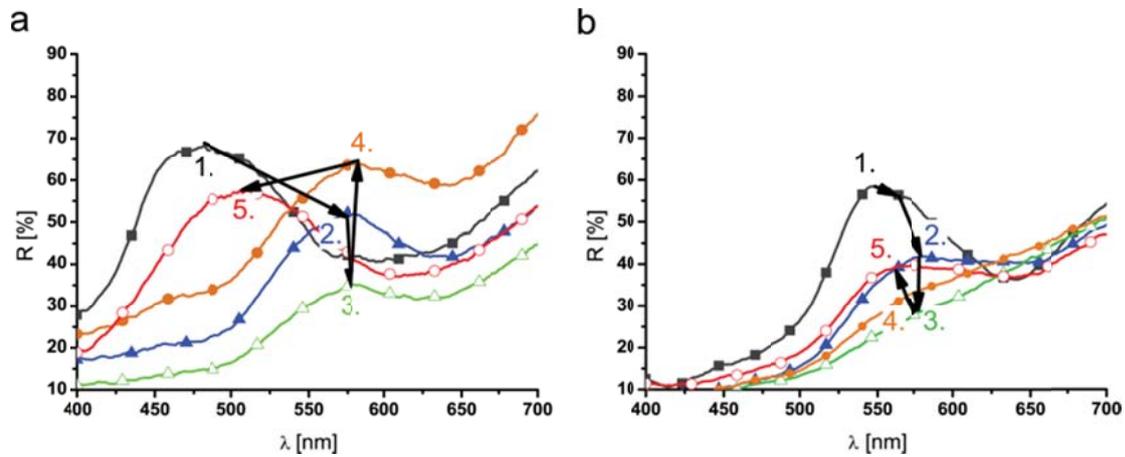

Figure 5: Spectral measurements on the blue/ventral (a) and green/dorsal (b) surfaces of the wings of the butterfly *Cyanophrys remus*. The numbered curves show the stationary reflectance curves in various stages: 1. dry status (23 °C); 2. condensation (10 °C); 3. freezing (-5 °C); 4. melting (0 °C); and 5. evaporation processes.

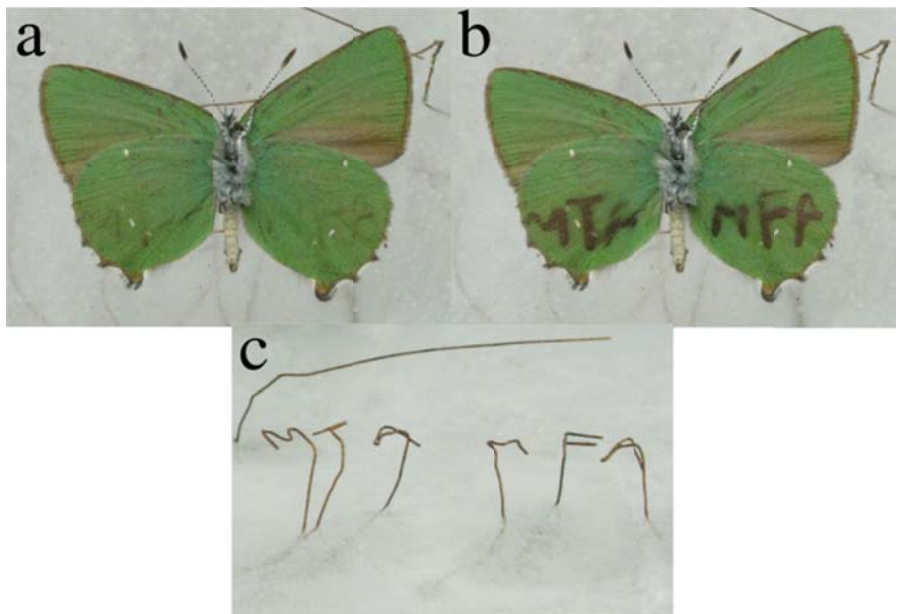

Figure 6: *Callophrys rubi* was placed onto cooled wires that were used to form characters. One end of the wires was frozen into ice, and the other end was used to form characters (c). The cold wires cool the surrounding area, where the condensation can begin. The images were collected 2-3 seconds (a) and 1 minute (b) after putting the samples onto the wires.

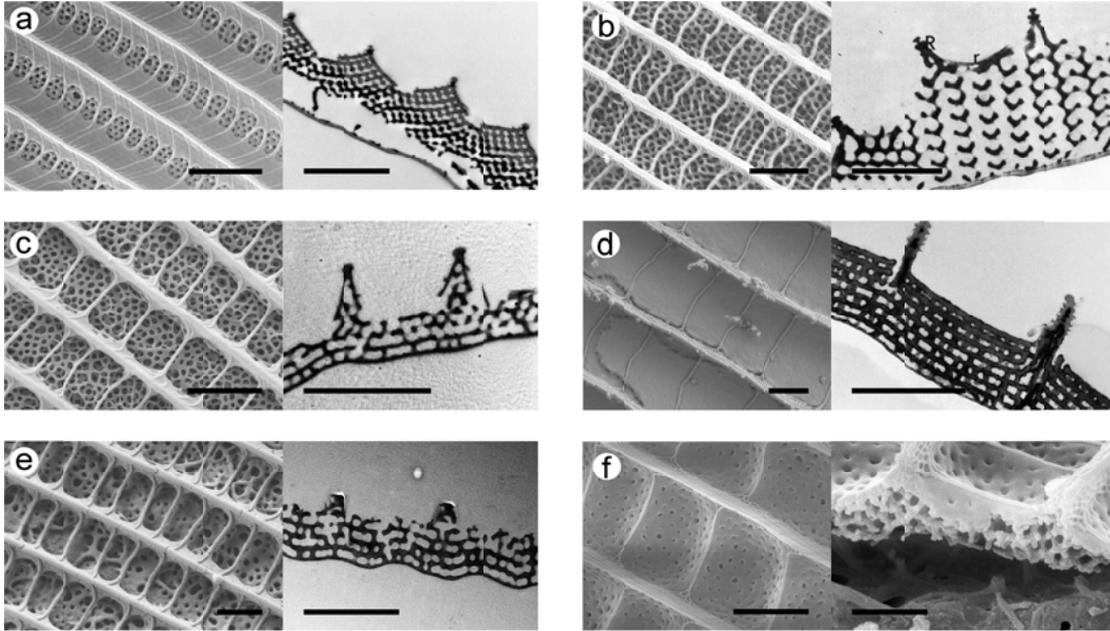

Figure 7: SEM/TEM images of the butterflies. Left: SEM; right: TEM, except for f), where both are SEM images. All scale bars are 2 μm. a) *Cyanophrys remus* (blue dorsal wing surface); b) *Callophrys rubi* (green dorsal wing surface, TEM image is reprinted from Ghiradella & Radigan 1976); c) *Polyommatus daphnis* (blue dorsal wing surface); d) *Chrysiridia ripheus* (blue-green ventral wing surface); e) *Albulina metallica* (blue dorsal wing surface); f) *Eterusia taiwana* (green dorsal wing surface).

**Electronic supplementary material**

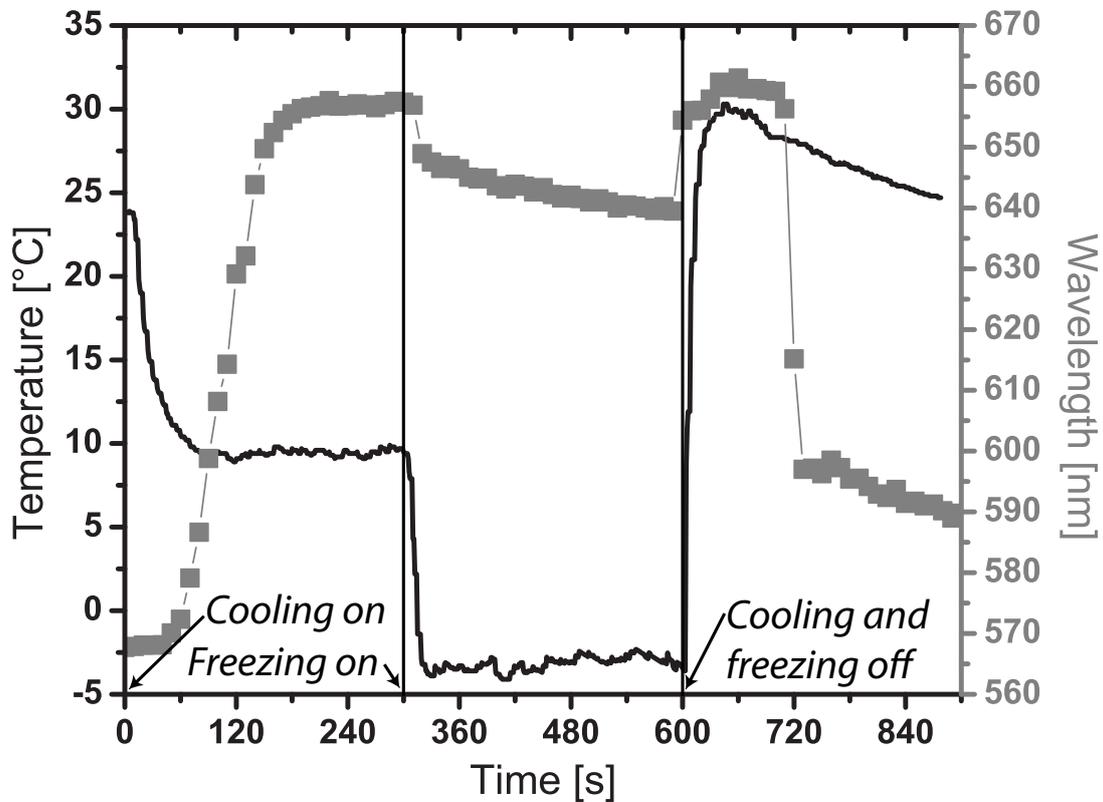

**F.S1:** Position of the reflection maximum (gray line with symbols) on the green ventral side of *A. metallica* during one whole cooling cycle (i.e., cooled below the dewpoint, then frozen and then the cooling apparatus was turned off). The temperature of the top surface of the Peltier cooler next to the wing (black line) was measured during one whole cooling cycle (i.e., condensation, freezing, melting and evaporation).

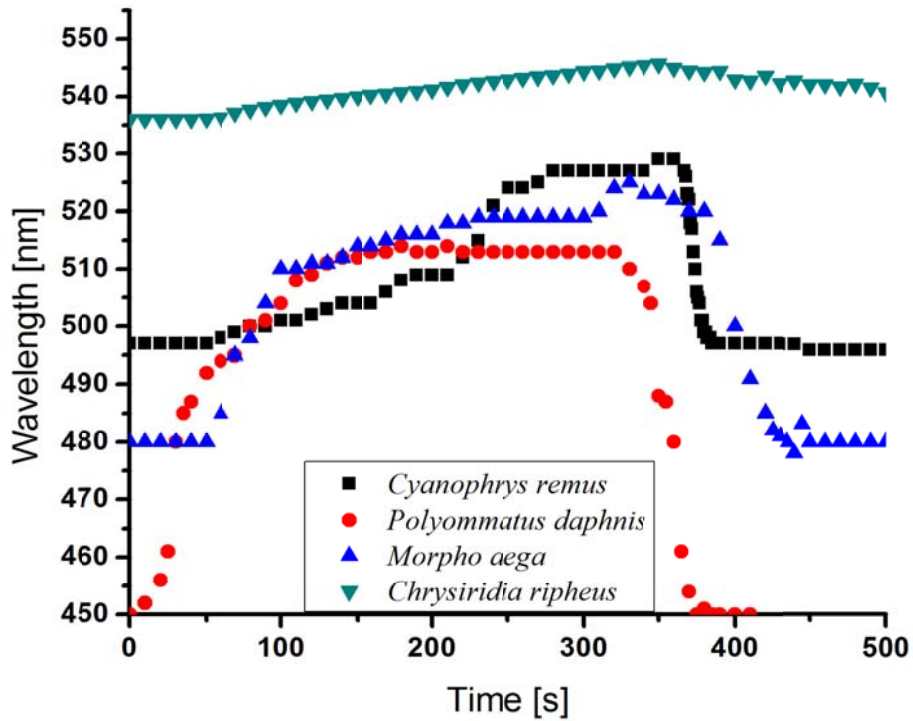

**F.S2:** Time dependence of the shifts in the reflection maxima in three blue species (*M. aega, P. daphnis, C. remus*) and in a species with closed photonic structures (green part of *C. ripheus*). The cooling apparatus was turned on at 50 sec, and it was turned off at 350 sec (gray lines). Relatively fast shifts of the reflection maxima could be observed for the three blue butterflies after the cooling apparatus was turned on, and much faster shifts were observed after it was turned off. The species with closed photonic structures exhibited very slow and small shifts in both cases, which could be interpreted as being due to the swelling of the structure and/or the very few open holes in the upper surface (epicuticle) of the nanostructure (see Fig. 7d).